\documentclass[aps,pre,twocolumn,amsmath,amsfonts,floatfix,superscriptaddress,longbibliography]{revtex4}
\usepackage{mathrsfs} \usepackage{amssymb} \usepackage{graphicx,color} \usepackage{bbm} \usepackage{multirow}
\usepackage{graphicx}
\usepackage{amsmath}
\usepackage{bbm} \usepackage{mathrsfs}
\usepackage{hyperref}
\usepackage[usenames,dvipsnames]{xcolor}
\usepackage{color}
\definecolor{darkblue}{rgb}{0,0,0.6}
\newcommand{\rs}{\rho_{\mathrm{swap}}}
\newcommand{\ns}{n_{\mathrm{swap}}}
\newcommand{\ta}{\tau_{\alpha}}
\newcommand{\tmd}{t_{\mathrm{MD}}}
\hypersetup{
bookmarksopen=true,
pdftitle="",
pdfauthor="",
pdftoolbar=false, 
pdfstartview={FitH},		
pdfmenubar=true,			
pdfhighlight=/O,			
colorlinks=true,			
urlcolor=darkblue,
citecolor=darkblue,		
linkcolor=darkblue}

\begin{document}

\title{Particle swap algorithms in discrete and continuous time}

\title{Efficient swap algorithms for molecular dynamics simulations of equilibrium supercooled liquids}

\author{Ludovic Berthier}
\affiliation{Laboratoire Charles Coulomb,
UMR 5221 CNRS-Universit\'e de Montpellier, Montpellier, France}

\author{Elijah Flenner}
\affiliation{Department of Chemistry, Colorado State University, Fort Collins, Colorado 80523, USA}

\author{Christopher J. Fullerton}
\affiliation{Laboratoire Charles Coulomb,
UMR 5221 CNRS-Universit\'e de Montpellier, Montpellier, France}
\affiliation{Department of Physiology, Anatomy and Genetics, University of Oxford, Oxford, United Kingdom}

\author{Camille Scalliet}

\author{Murari Singh}

\affiliation{Laboratoire Charles Coulomb,
UMR 5221 CNRS-Universit\'e de Montpellier, Montpellier, France}

\begin{abstract}
It was recently demonstrated that a simple Monte Carlo (MC) algorithm involving the swap of particle pairs dramatically accelerates the equilibrium sampling of simulated supercooled liquids. We propose two numerical schemes integrating the efficiency of particle swaps into equilibrium molecular dynamics (MD) simulations. We first develop a hybrid MD/MC scheme combining molecular dynamics with the original swap Monte Carlo. We implement this hybrid method in LAMMPS, a software package employed by a large community of users. Secondly, we define a continuous time version of the swap algorithm where both the positions and diameters of the particles evolve via Hamilton's equations of motion. For both algorithms, we discuss in detail various technical issues as well as the optimisation of simulation parameters. We compare the numerical efficiency of all available swap algorithms and discuss their relative merits.
\end{abstract}

\maketitle

\section{Introduction}

Simulations are a useful tool to understand the equilibrium properties of supercooled liquids approaching a glass transition~\cite{BB11}. They offer microscopic insight into static and dynamic properties for well-defined and usually quite simple model systems. A major obstacle in this approach is the difficulty of simulating large enough timescales in order to get closer to experimentally-relevant studied materials. Recently, the gap between simulated timescales and experimental ones has been closed~\cite{BCNO15,NBC17} using a simple Monte Carlo (MC) scheme, where ordinary translational moves of the particles (that mimic the physical dynamics) are complemented by the swap of unlike particle pairs~\cite{oldfrenkel,GP01}. The method is thus broadly applicable to models composed of discrete or continuous mixtures of distinct particles. In practice, this encompasses virtually all types of glass-formers~\cite{BB11}.

While having many advantages, such as simplicity and efficiency, Monte Carlo is not necessarily the most commonly used simulation technique for supercooled liquids. Molecular dynamics (MD) techniques are often preferred, because the microscopic dynamics is closer to that of real molecular fluids~\cite{AllenTildesley}. For colloidal particles, Brownian dynamics may be more adapted. For several models of supercooled liquids, the equivalence of all these different microscopic dynamics to those of MC simulations is fully established~\cite{berthierkobLJ,berthierSILICA,BBBKMR07a,BBBKMR07b}. This implies in particular that the chosen numerical method to simulate supercooled liquids is essentially one of personal convenience. Since particle swaps were first introduced in the context of MC studies~\cite{GP01,BCNO15}, it is natural to ask whether it is possible to extend the method in the more general context of MD simulations.

One potential advantage of MD simulations is that they are easier to parallelize than MC techniques, mainly because the positions of the particles are all updated simultaneously in MD rather than sequentially in MC. This makes simulating large systems prohibitively slow when using MC simulations.
Although spatial correlations remain relatively modest in supercooled liquids~\cite{BB11,BBBCS11}, larger and larger systems are required to analyse the deeply supercooled states that swap MC simulations can now potentially access~\cite{ceiling,ceiling2d,chris,yielding}. It is therefore natural to ask if it is possible to introduce swap moves into molecular dynamics, a simulation method which is readily parallelized.

As a first step we consider a hybrid MC/MD simulation scheme where blocks of swap Monte Carlo moves are inserted at regular time intervals into a conventional molecular dynamics simulation. This method has already been used long ago~\cite{oldfrenkel}, in a different context. We show how to best tune the parameters of this hybrid technique to obtain maximum efficiency, and carefully discuss the numerical efficiency of the technique. We implement the method into the LAMMPS open software for MD simulations~\cite{plimpton}. In a second effort, we implement a continuous time version of the swap MC into a generalized MD scheme where both the positions and the diameters of the particles obey Newton's equations of motion for a suitably defined Hamiltonian. This second scheme bears similarities with semi-grand MC techniques for polydisperse fluids~\cite{glandt,oldfrenkel,lernerprx}. When properly optimised, we find that all three simulations techniques provide essentially the same (potentially very large) speedup over conventional MD and MC techniques, so that again the choice of one swap algorithm over another is mostly one of personal convenience.

This article is organised in the following way.
In Sec.~\ref{sec:hybrid}, we discuss the hybrid MC/MD method: numerical scheme, details of the studied models, thermalisation speedup.  These results are of general interest and do not depend on the specific implementation of the algorithm. We then present the computational efficiency obtained with our implementation of the method in the LAMMPS package. In Sec.~\ref{sec:continuous}, we present the continuous time version of the swap algorithm. The relative merits of all swap algorithms are discussed in Sec.~\ref{discussion}.

\section{Hybrid MC/MD method}
\label{sec:hybrid}

\subsection{Microscopic model}
\label{subsec:model}

We study numerically a system composed of $N$ size polydisperse particles of identical masses $m$ in a cubic box of linear size $L$ with periodic boundary conditions. The system is defined by the 3$N$ particle position coordinates ${\bf r}^N = \{{\bf r}_1,{\bf r}_2, \ldots ,{\bf r}_N\}$, and the particle sizes $\sigma ^N= \{\sigma_1, \sigma_2, \ldots ,\sigma_N\}$.

Two particles $i \neq j$ at a distance $r_{ij} = |{\bf r}_i - {\bf r}_j|$ interact only if $r_{ij} < 1.25 \sigma_{ij} $. We model the interactions between particles via a soft repulsive pair potential
\begin{equation}
\label{eq:pot}
\begin{split}
u(r_{ij},\sigma_{ij}) / \epsilon &= \left(\frac{\sigma_{ij}}{r_{ij}}\right)^{12} +  F(r_{ij},\sigma_{ij})  ~~ , \\
\text{where }F(r_{ij},\sigma_{ij})  &= c_0  + c_2 \left(\frac{r_{ij}}{\sigma_{ij}}\right)^{2} + c_4 \left(\frac{r_{ij}}{\sigma_{ij}}\right)^{4} ~~
\end{split}
\end{equation}
is a function that smooths the potential at the cutoff distance $1.25\sigma_{ij}$. The coefficients $c_0$, $c_2$, and $c_4$
ensure the continuity of the potential up to the
second derivative at the cutoff. The total potential energy of the system is $U({\bf{r}}^N,\sigma^N) = \sum_{i<j} u(r_{ij},\sigma_{ij})$.
In order to obtain a good glass-forming model, we study a continuously polydisperse system. The particle diameters follow the distribution $P(\sigma_{m} \leq \sigma \leq \sigma_{M}) = A/\sigma^{3}$, where $A$ is a normalizing constant, $\sigma_{m} = 0.73$ and $\sigma_{M} = 1.62$.
To ensure the structural stability of the polydisperse mixture, we employ a nonadditive interaction rule for the cross diameters $\sigma_{ij}= \frac{\sigma_i+\sigma_j}{2}(1 - 0.2 |\sigma_i-\sigma_j| )$, following previous work~\cite{NBC17}. Lengths and times are respectively expressed in units of $\overline{\sigma} = \int  \sigma P(\sigma) d\sigma$ and $\sqrt{\epsilon / m \overline{\sigma}^2}$. In the following, we present results for this model at number density $\rho = N/L^3 = 1$, mostly for $N = 1500$. In Sec.~\ref{subsec:parallel}, we study systems with larger sizes to check the scalability of the algorithm with system size.

\subsection{Hybrid scheme}
\label{subsec:scheme}

\begin{figure}
	\begin{center}
		\includegraphics[width=7.5cm]{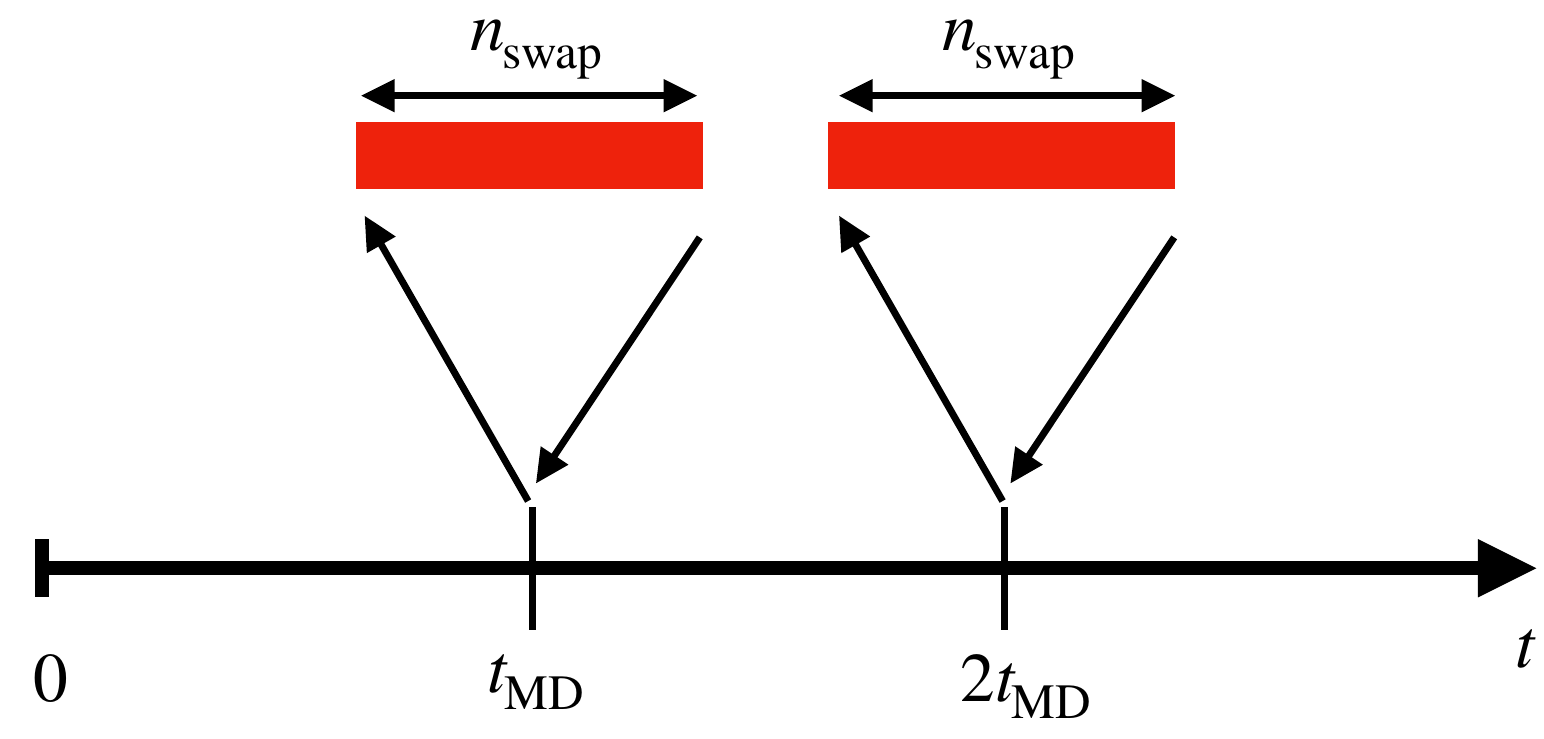}
		\caption{The hybrid scheme consists of a regular succession of blocks of molecular dynamics simulations and blocks of particle-swap Monte Carlo steps. Every $t_{\mathrm{MD}}$, the molecular dynamics is paused and $n_{\mathrm{swap}}$ swap Monte Carlo steps are performed, which are not counted in the total elapsed MD time.}
		\label{fig:hybrid_scheme}
	\end{center}
\end{figure}

We introduce the hybrid scheme used to simulate the glass-forming model presented in Sec.~\ref{subsec:model}. The method consists of alternating between ordinary molecular dynamics simulation sequences during which the particle positions evolve with a fixed particle size, and particle-swap Monte Carlo sequences during which the particles exchange their sizes at fixed positions. The hybrid method is illustrated in the schematic diagram of Fig.~\ref{fig:hybrid_scheme}.

The trajectories of particles in the MD blocks are generated in the canonical ensemble ($NVT$) by integrating Nos\'e-Hoover chain equations of motion~\cite{Nose85,Nosefail01,NHC92,NHC96}. We use a chain of thermostats of length three. The time integration of the equations of motion is performed by a time-reversible measure-preserving Verlet algorithm, with a time discretization  $dt =0.01$~\cite{Tuckerman06}. The damping parameter associated to the heat bath variables is equal to $1$. The particles positions and velocities are evolved during sequences of duration $t_{\mathrm{MD}}$. This defines the MD blocks. At the end of each MD block, the time is paused, and the particle positions and velocities are frozen. A series of particle-swap Monte Carlo moves are then performed, and this defines a swap Monte Carlo (SMC) block. These blocks are composed of $N_{\mathrm{swap}}$ attempted elementary swap moves. During each elementary move, two particles are chosen randomly and the exchange of their diameters is accepted or rejected based on the Metropolis criterion. The swap moves preserve detailed balance and guarantee an equilibrium sampling of phase space in the $NVT$ ensemble. The duration of the swap Monte Carlo blocks is defined in a system-size independent way through $n_{\mathrm{swap}} = N_{\mathrm{swap}}/N$.

We combine the parameters $\tmd$ and $n_{\mathrm{swap}}$ together as
\begin{equation}
\rs = \frac{\ns}{\tmd},
\end{equation}
which represents the density of particle-swap Monte Carlo moves per particle and unit MD time. In the following, we will study how the parameters $\tmd$ and $\ns$ affect the efficiency of the algorithm. In particular, we will study the competition between the thermalisation speedup offered by the swap moves and the additional CPU time entailed by the addition of swap MC blocks.

We have implemented the hybrid scheme into the LAMMPS open software, because it is widely used and versatile. We have already used this method to study other model systems, such as Lennard-Jones and its truncated Weeks-Chandler-Andersen version~\cite{wca3d,wca3dd,SBZ17}. Very deeply supercooled states have been successfully obtained for all these models. Details about how size polydispersity is handled and other LAMMPS-specific details are presented in Appendix \ref{sec:lammps}.

\subsection{Proper sampling of the canonical ensemble}
\label{subsec:thermo}

In the previous section, we presented the hybrid scheme as a succession of molecular dynamics and swap Monte Carlo blocks. Inside each block, the dynamics (MD or MC) is carefully designed to sample the canonical $NVT$ ensemble. However, given the different nature of both algorithms, we need to ensure that the combination of both methods continues this equilibrium sampling.

In the hybrid method, the MD simulation is regularly interrupted to perform particle-swap moves. In the MD blocks, both the potential and kinetic energies fluctuate. By contrast, the SMC blocks only affect the potential energy of the system, since only the diameters of particles are changed, at fixed positions and velocities. As a new MD block starts, the particles have the same positions and velocities as in the previous MD block, but they now possess a different potential energy. It takes a short but finite time (of the order $t \sim 0.1$) for the kinetic energy to relax after the SMC block has been performed. As a result, MD blocks cannot be made arbitrarily short. When hybrid simulations are run with $\tmd < 0.1$, we have found that the system heats. In the limit $\tmd = dt$, which amounts to alternating MC and MD at each integration step, the kinetic energy is up to $3\%$ higher than the imposed temperature and the Nos\'e-Hoover thermostat does not work properly.

However, when the hybrid simulations are run with $\tmd > 0.1$, the probability distributions of the potential and kinetic energies follow the canonical ones, and coincide with those obtained from standard $NVT$ simulations without the SMC blocks.

\subsection{Equilibration speedup}
\label{subsec:physeff}

In this section, we study the equilibrium dynamics of the model presented in Sec.~\ref{subsec:model} simulated with the hybrid method.

We first run equilibration simulations during which we monitor the evolution of the potential energy $U$ and the structure factor of the liquid, to detect aging effects and potential instabilities of the homogeneous fluid. When equilibrium is reached, we compute the self-part of the intermediate scattering function
\begin{equation}
\label{eq:self}
F_s(k,t)=  \left\langle
\frac{1}{N}\sum_j e^{i\textbf{k}\cdot[\textbf{r}_j(t)-\textbf{r}_j(0)]} \right\rangle~~.
\end{equation}
We spherically average over wavevectors of magnitude $k = 7.0$, which corresponds to the first diffraction peak in the static structure factor of the
liquid. The brackets indicate averages over independent equilibrated configurations. Following common practice, we define the structural relaxation time of the liquid $\ta$ as the time at which $F_s(k,\tau_\alpha)=e^{-1}$.

\begin{figure}[tb]
\includegraphics[scale = 1]{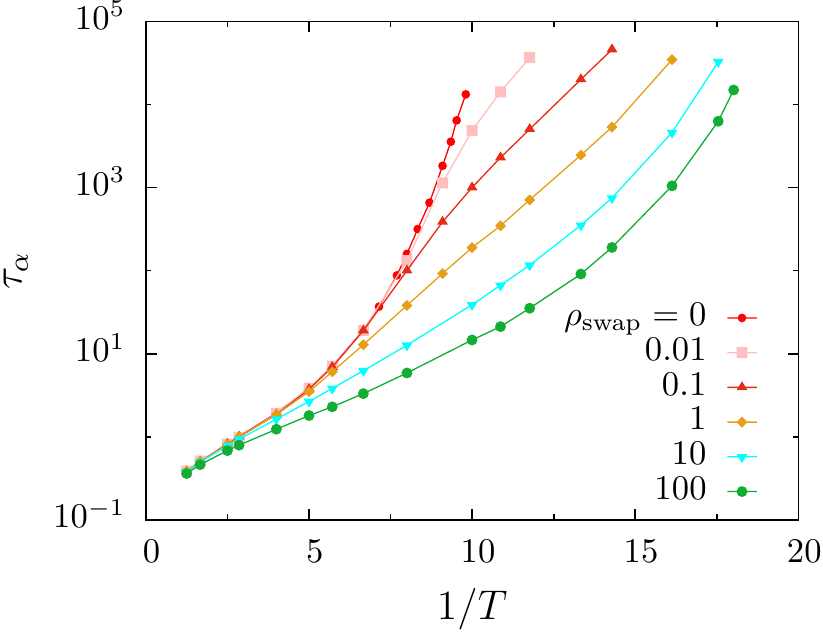}
\caption{Evolution of the equilibration time $\ta$ with inverse temperature $1/T$ in hybrid MD/MC simulations of the three-dimensional soft polydisperse model of Sec.~\ref{subsec:model} at number density $1$. The swap density $\rs$ is varied between $\rs = 0$ (ordinary molecular dynamics) and $\rs = 100$.  The dynamics at intermediate $\rs$ smoothly interpolates between these two limits.}
\label{fig:angell_rhos}
\end{figure}

We study the influence of gradually adding SMC blocks to standard MD simulations. We compute the equilibrium relaxation time of the liquid varying temperature and swap density $\rs$ and report the results in Fig.~\ref{fig:angell_rhos}. For $\rs \geq 0.1$, we use $\tmd = 0.1$, and different lengths $\ns$ for swap blocks. To access the lowest density of swap $\rs = 0.01$, we use instead $\tmd = 1$. The resulting swap density $\rs$ then varies from $0.01$ to $100$. Conventional molecular dynamics simulations correspond to $\rs = 0$.

For standard molecular dynamics, the relaxation time of the liquid increases sharply as temperature decreases. We evaluate empirically $T_{MCT} \approx 0.1$ through a mode-coupling theory power-law fit to the relaxation time data~\cite{Go99}. In practice, the numerical study of equilibrium supercooled liquids with molecular dynamics simulations are confined to temperatures above $T_{MCT}$, as the relaxation time near $T_{MCT}$ corresponds typically to the maximal computer time accessed in conventional MD. As such the mode-coupling temperature crossover $T_{MCT}$ is a useful temperature scale in the context of computer simulation studies of supercooled liquids.

The situation changes gradually as $\rs$ increases. For low values $\rs = 0.01 - 0.1$, the relaxation time of the system at high temperature is equal to that of the standard MD simulation. In this regime, the dynamics is not slow enough to be affected by the addition of a small number of swap moves. Around $T_{MCT}$ and below, the addition of short SMC blocks becomes key to observing the liquid relax in numerically accessible timescales. Strikingly, equilibrium is easily reached at $T< T_{MCT}$ with a density of swap as small as $\rs = 0.01$, i.e., when only $1\%$ of the $N$ particles are swapped per unit time. As the swap density $\rs$ increases, the relaxation time departs more strongly from the one obtained with pure MD simulations.

For large swap densities such as $\rs = 100$, the dynamics are affected so much that the high-temperature Arrhenius behavior of the normal dynamics now persists almost down to $T_{MCT}$, before eventually increasing with a super-Arrhenius law at lower temperatures. We find that the hybrid method can achieve thermalization of supercooled liquids down to $0.6T_{MCT}$. We point out that Fig.~\ref{fig:angell_rhos} resembles results obtained with the swap Monte Carlo method, in which the role of $\rs$ was played by the probability, $p$, to perform a particle-swap move over a translational move~\cite{BBBT18}. This suggests a close correspondence between $\rs$ in the hybrid method and $p$ in the original swap MC method, which we explore further in Sec.~\ref{subsec:smchybrid}.

\begin{figure}
\includegraphics[scale = 1]{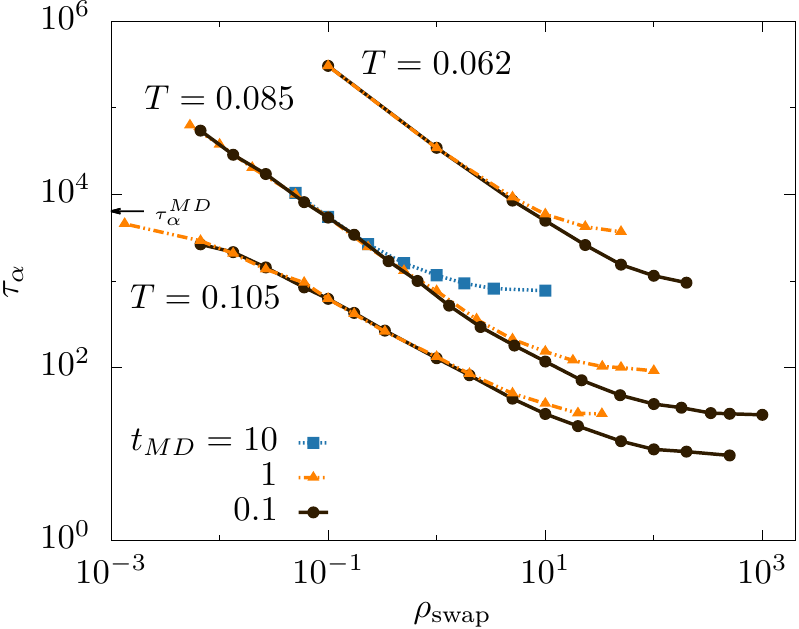}
\caption{Relaxation time as a function of $\rs$ for three selected temperatures: $T = 0.105, 0.085, 0.062$. At fixed temperature, each data set corresponds to a given duration of the MD blocks, $\tmd = 0.1, 1, 10$, and each single point to one value of $\ns$. The arrow labeled $\ta^{MD}$ indicates the relaxation time of the liquid at $T = 0.105$ for ordinary MD simulations; this time cannot be directly measured for the two lowest temperatures.}
\label{fig:rsvar}
\end{figure}

In order to optimize the hybrid method, we investigate the separate influence of the parameters $\ns$ and $\tmd$ on the thermalization speedup of supercooled liquids. We select three temperatures of interest, one above $T_{MCT}$ and two below: $T = 0.105$, $0.085$, and $0.062$. For each temperature, we report the relaxation time $\ta$ measured with the hybrid simulations at different $\ns$ and $\tmd$ in Fig.~\ref{fig:rsvar}. Each data set corresponds to a fixed duration $\tmd$ of the MD blocks, so that increasing $\rs$ corresponds to increasing the duration $\ns$ of the SMC blocks.

The qualitative evolution of the relaxation time with swap density is similar at all temperatures. Starting from the limit $\rs = 0$ (conventional MD simulations), $\ta$ decreases with the swap density, roughly as $\ta \sim 1/\rs$~\cite{NBC17}. At $T = 0.085, 0.062$, this behaviour is observed over a few decades of $\rs$. At large $\rs$, all curves saturate to a plateau value: increasing the length of SMC blocks at fixed interval $\tmd$ does not speedup further the dynamics.
Indeed, as $\ns$ increases more particle-swap attempts are performed in a SMC block. But since the particles' positions are frozen during such SMC blocks, the saturation of $\ta$ reflects the thermalization of the particles' diameters within a frozen configuration. At a given temperature, the plateau value at larger $\rs$ depends on $\tmd$: the longer $\tmd$, the higher $\ta$ is at the plateau. Since molecular dynamics is inefficient at relaxing the structure of the liquid in this temperature regime, longer MD blocks do not help speedup the structural relaxation. We emphasise that MD blocks are nonetheless essential to the hybrid method, since swap moves and particle displacements work hand in hand to decorrelate the structure of the liquid~\cite{NBC17}.

The optimal value $\tmd = 0.1$ emerges from optimizing the physical efficiency (see Fig.~\ref{fig:rsvar}) of the hybrid simulation which requires a small $\tmd$ value, together with the constraint that $\tmd$ must be large enough for a proper sampling of the canonical ensemble (see Sec.~\ref{subsec:thermo}).

\subsection{Efficiency of the hybrid method on a single CPU}
\label{subsec:serial}

\begin{figure}
\includegraphics[scale = 1]{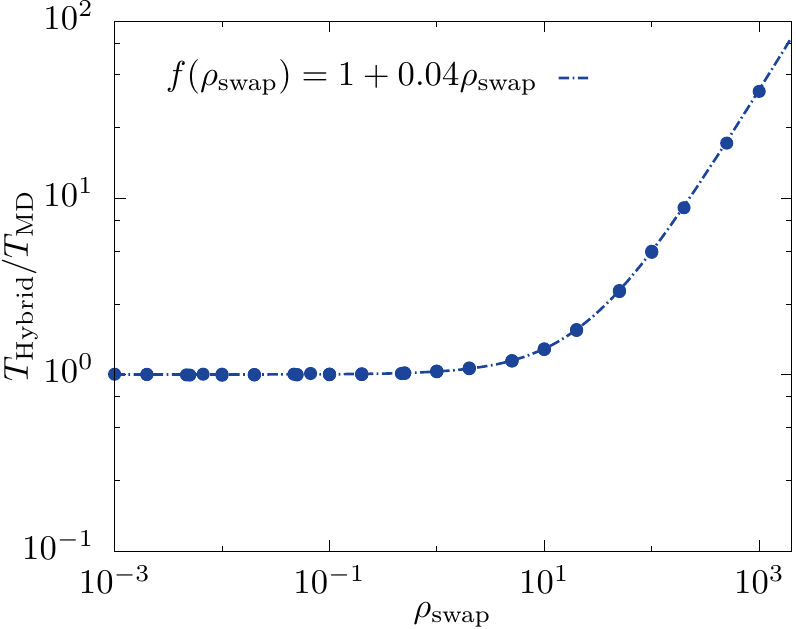}
\caption{Ratio of the CPU time $T_{\mathrm{Hybrid}}$ of Hybrid simulations compared to the CPU time of standard molecular dynamics $T_{\mathrm{MD}}$, both running for the same total MD length. The trivial limits of $T_{\mathrm{Hybrid}}/T_{\mathrm{MD}}$ at small and large $\rs$ are well captured by a  simple empirical fitting function shown with a dashed line.}
\label{fig:CPUserial}
\end{figure}

In this section, we focus on the efficiency of the hybrid method executed on a single CPU. More specifically, we are interested in quantifying the competition between the added CPU cost due to increasing the number of swap moves and the speedup in thermalisation offered by the swap moves observed in Fig.~\ref{fig:angell_rhos}. Such results, therefore, will be a combination of the physical efficiency of the algorithm, the efficiency of our implementation of the hybrid method, and the hardware that we run it on. However, our discussion is generic and should be useful to anyone willing to employ the hybrid method. We present results obtained with our implementation of the hybrid method in the LAMMPS package. We expect these results to be broadly applicable, as there is little flexibility in implementing such a serial program, apart from well-known optimizations~\cite{AllenTildesley}.

To characterize the influence of $\rs$ on the CPU time in the hybrid method, we measure the time in seconds to run hybrid simulations which last the same total MD time, using different combinations of $\ns$ and $\tmd$. The computational time should of course not depend on how the MD and SMC blocks are distributed, but rather on the total duration of each type of blocks. We therefore report times as a function of $\rs$ in Fig.~\ref{fig:CPUserial}. As expected, we see that all points collapse on a single master curve, confirming that the CPU time indeed depends on $\rs$ only. In the range $\rs = 0-10$, the CPU time of the hybrid method is dominated by that of the MD blocks. Around $\rs = 10$, the CPU time becomes dominated by particle-swaps, and eventually grows linearly with $\rs$ with a slope controlled by the CPU cost of an individual swap move. This simple dependence of the CPU time with $\rs$ is well captured by a fitting function $f(\rs) = 1 + 0.04 \rs$ (dashed line in Fig.~\ref{fig:CPUserial}), which captures these two limits.

To finally determine the optimal parameters of the hybrid method, we combine the dynamical gain shown in Fig.~\ref{fig:rsvar} and the computational cost discussed in Fig.~\ref{fig:CPUserial}. The product of both quantities quantifies the time needed to achieve a given number of MD steps in units of the relaxation time of the system. In other words, this quantifies how long (in CPU time) it takes to equilibrate the system at a particular state point. This quantity should be minimal for the hybrid method to be the most efficient.

\begin{figure}
\includegraphics[scale = 1]{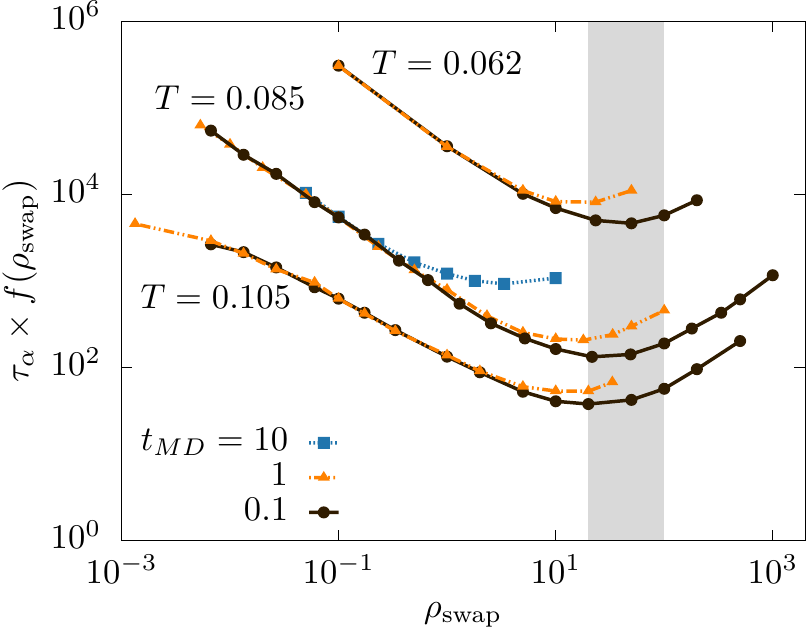}
\caption{The product of the measured relaxation time $\ta$ with the computational cost $f(\rs)$ of increasing the swap density in the hybrid method presents a minimum at all temperature. The hybrid method is the most efficient with the parameters yielding a minimum in the curves. The best trade-off between physical speedup and CPU cost is reached for $\ns = 2-10, \tmd= 0.1$, as highlighted by the shaded region.
\label{fig:toteff1cpuu}}
\end{figure}

The numerical results are shown in Fig.~\ref{fig:toteff1cpuu}. All the curves shown in this figure present a minimum for a given value of $\rs$, and the location and value of this minimum both depend on $\tmd$. For a given temperature, the global minimum occurs for $\tmd=0.1$. The location of the minimum varies over a narrow range of $\rs$, slightly shifting to higher $\rs$ at lower temperatures. This range is highlighted by the shaded region in Fig.~\ref{fig:toteff1cpuu}, and corresponds to $\rs = 20-100$ and thus to
$\ns = 2-10$. This represents the best trade-off between the speedup offered by increasing the number of swap moves, and the added CPU cost of performing these moves.

\subsection{Comparison between the hybrid and swap MC methods}
\label{subsec:smchybrid}

We now compare the efficiency and physical dynamics obtained in both the hybrid and the swap MC methods. Both methods have their own set of optimised parameters. For the hybrid method, $\tmd$ and $\ns$ must be tuned, whereas for swap MC one must adjust the relative probability $p$ to attempt a particle-swap move instead of a translational move. Within the MC approach the typical size of the translational moves must also be adjusted~\cite{berthierkobLJ}. In order to compare the two methods, we present results for simulations run with the optimal parameters in each case. The optimal efficiency of the swap MC algorithm is reached around $p = 0.2$~\cite{NBC17}. In this section, we consider hybrid simulations with $\ns = 10, \tmd = 0.1$.

\begin{figure}
\includegraphics[scale = 1]{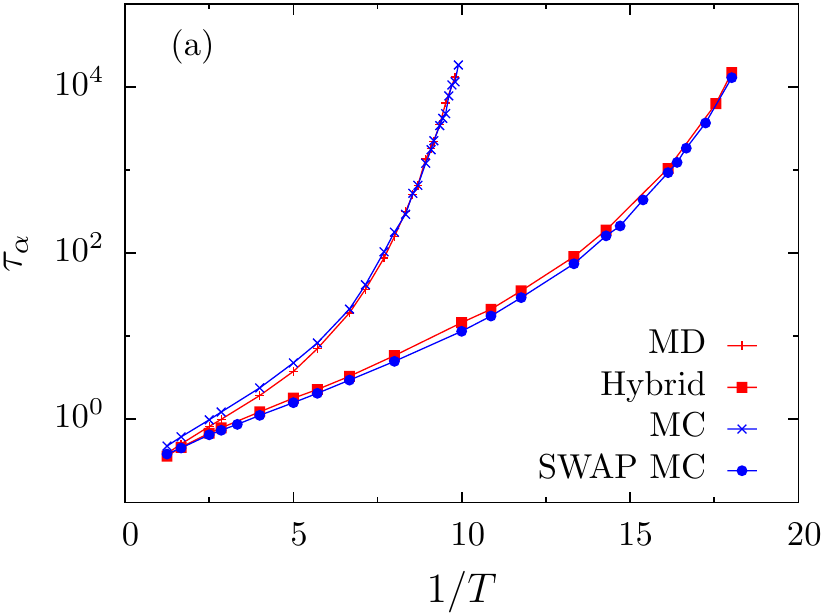}
\includegraphics[scale = 1]{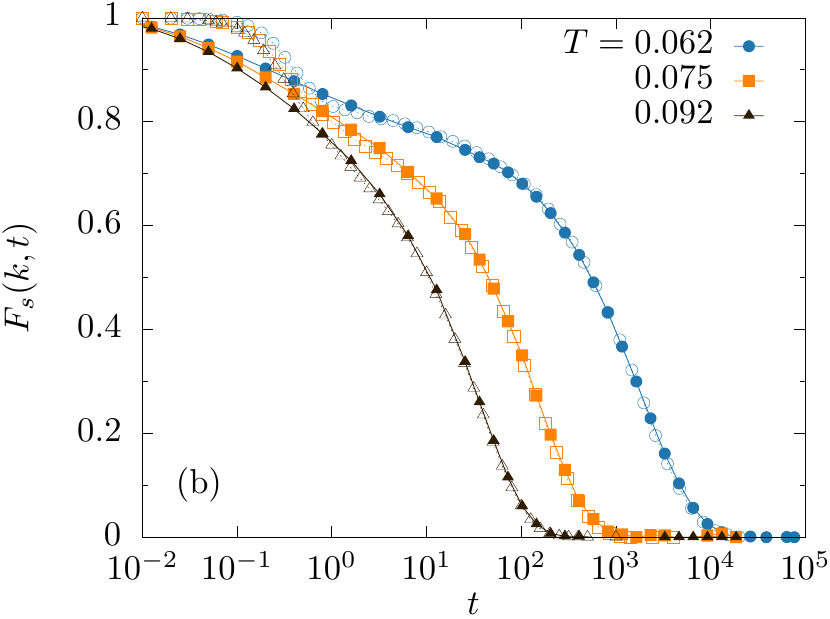}
\caption{Comparison between the hybrid ($\ns = 10,~\tmd = 0.1$) and swap MC algorithms ($p = 0.2$).
(a) Equilibrium relaxation times of the liquid $\ta$ as a function of the inverse temperature in hybrid and swap MC methods, as well as in standard MD and MC simulations. Relaxation times for hybrid and MD methods are in MD units. For swap and conventional MC, we convert 1 MC step into $a=3.2$ MD steps.
(b) Self-intermediate scattering function $F_s(k,t)$ measured in swap MC (close symbols) and hybrid simulations (open symbols) at $T = 0.062,0.075,0.092$ using the same time units as in (a).
These data demonstrate the full equivalence between swap MC and hybrid simulations, which offer the same equilibration speedup over conventional MC and MD methods.}
\label{fig:MCvsMD}
\end{figure}

To compare MD and MC methods, we need to employ a dictionary between timescales, which correspond to very different processes in both approaches.
To this end, we first measure the relaxation time of supercooled liquids measured in standard MC and MD simulations, i.e. with no swap moves at all. These are respectively expressed in numbers of MC steps and MD time. As found before in a different system~\cite{berthierkobLJ}, we observe that the structural relaxation time in both dynamics follows a similar temperature dependence, see Fig.~\ref{fig:MCvsMD}a. Rescaling the MC curve on top of the MD curve, we find that $t=1$ in MD units corresponds to $t \approx 320$ MC steps.
Using a time discretisation $dt=0.01$, this implies that 1 MD step corresponds roughly to $a=3.2$ MC steps, a conversion similar to the one found for a Lennard-Jones model~\cite{berthierkobLJ}.

This conversion factor allows us to convert the simulation parameters used in optimal hybrid simulations, $\ns = 10, \tmd = 0.1$, into an equivalent probability of performing swap moves: $p^{\rm equiv} = (\ns/a) / (\ns/a + \tmd/dt) \approx 0.238$, which is indeed very close to the optimal $p \approx 0.2$ determined in Ref.~\cite{NBC17}.

In Fig.~\ref{fig:MCvsMD}b, we show self-intermediate scattering functions $F_s(k,t)$ measured in hybrid and swap MC simulations at three temperatures below $T_{MCT}$. We have converted Monte Carlo steps in MD units using the above conversion factor. We see that the equivalence between MC and MD dynamics discussed before for conventional simulations~\cite{berthierkobLJ,berthierSILICA} now extends to swap algorithms. Apart from small differences at short times, the decay of time correlation functions using swap MC and the hybrid methods are very similar.

We obtain the relaxation time for these two swap dynamics and present the results in Fig.~\ref{fig:MCvsMD}a along with the results for the ordinary dynamics. It is clear from this figure that the relaxation times of the swap MC and hybrid methods are again equivalent. We conclude that the hybrid method is able to speedup the equilibration of supercooled liquids with an efficiency comparable to the one of the original swap MC algorithm. Given the above conversion factor of order unity between MC and MD steps, we finally conclude that both methods give an equivalent equilibration speedup at an equivalent CPU cost.

We have shown that the hybrid method MC/MD is as powerful as the swap Monte Carlo algorithm when it comes to generating computer supercooled liquids at temperatures lower than the laboratory glass transition. The implementation in the LAMMPS package that we propose should in addition makes this algorithm a very powerful and versatile tool accessible to the glass community.

\subsection{Efficiency of the hybrid method in parallel}

\label{subsec:parallel}

\begin{figure}
\includegraphics[scale = 1]{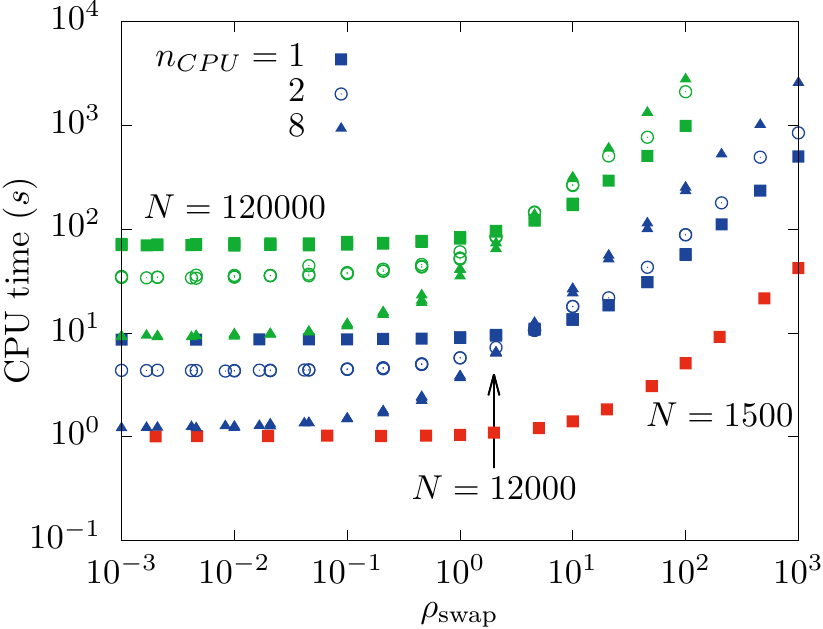}
\caption{CPU time (in seconds) as a function of the swap density $\rs$ for hybrid simulations of systems composed of $N = 1500$ (red), $12000$ (blue), $120000$ (green) particles, running on one (square), two (circle) or eight (triangle) processors. All the simulations run for a total time of $10$ (in MD units).}
\label{fig:cpuparallel}
\end{figure}

In essence, the hybrid method converts the translational moves of the original swap MC algorithm into MD integration steps, while keeping the much less frequent swap moves unchanged.  An important difference between translational MC steps and MD steps is that the former need to be performed sequentially, which makes MC intrinsically difficult to parallelise. Existing solutions to this problem only become advantageous for extremely large system sizes~\cite{ANDERSON201327}. Converting MC steps to MD steps in the hybrid method thus makes it possible to easily parallelise the translational part of the swap algorithm. This is an important objective of the present work.

The LAMMPS package, a ``Massively Parallel Simulator'', provides a good starting point to implement the hybrid scheme on several CPUs. In LAMMPS, the molecular dynamics is already well optimized to run on several processors. It is possible to parallelize MD simulations because the algorithm is deterministic, so each processor can be in charge of a subset of the total system without having to perform time-costly inter-processor communications frequently. To work within the existing framework of LAMMPS, some inter-processor communication is necessary during the SMC blocks. We now determine how much of an effect this has on the efficiency of the hybrid method when run in parallel.

We simulate at temperature $T=0.062$ systems composed of $N = 1500$, $12000$, $120000$ particles. The simulations have been run on one, two and eight processors. For a given system size and number of CPUs, we run simulations at different values of $\ns$ and $\tmd$. All the simulations are run for the same total MD length. In Fig.~\ref{fig:cpuparallel} we report the CPU time in seconds for this large set of hybrid simulations.

We observe two regimes in this figure. At low density of swap moves, $\rs < 1$, the CPU time of the simulation is dominated by the MD blocks. In this regime, the CPU time depends essentially on the number of particles per processor. For example, $N = 12000$ particles on 8 CPUs takes about the same CPU time as $N = 1500$ particles on one processor. In this regime of modest swap density, the hybrid algorithm allows us to efficiently simulate  very large systems by using more than one processor. In other words, we benefit from the optimal parallelisation offered by the MD algorithm, as implemented in LAMMPS. In this regime of system sizes, no such improvement would be gained for the original swap MC method.

At larger $\rs$, the CPU time becomes dominated by the SMC blocks, and it increases linearly with $\rs$, as found already in Fig.~\ref{fig:CPUserial}. The relative position of the curves corresponding to different numbers of CPUs is inverted compared to the low $\rs$ regime. In other words, running simulations on more processors does not decrease the CPU time of a simulation, bur rather increases it.

The difficulty in parallelising Monte Carlo algorithms is well-known and intrinsic to their stochastic sequential nature. In LAMMPS, information about particles in different parts of the box is stored on different processors. Adding SMC moves to LAMMPS therefore means that processors need to exchange information during most swap moves. These communications are time-consuming, and more frequent than during parallelised MD simulations. This means that swap moves in this implementation are less efficient in parallel than they are in serial. The CPU time in this regime of $\rs$ is completely dominated by these inter-processors communications, hence the increase in CPU time when running on more processors. More details are given in Appendix~\ref{sec:lammps}.

There is a crossover between the two regimes discussed above, at which the CPU time is the same for a given system size, regardless the number of processors. This crossover occurs at a value around $\rs \sim 10$, which tends to decrease as system size increases.

In order to get the global efficiency of the hybrid method in parallel, we have reproduced the analysis done in Fig.~\ref{fig:toteff1cpuu}. We multiply the physical relaxation time by the CPU time for simulations run in parallel. The optimal parameters for the hybrid method in serial are around $\rs = 20-100$ (see Fig.~\ref{fig:toteff1cpuu}) but this corresponds to the swap-dominated regime. As a result, the global efficiency of the hybrid method does not increase with the number of processors. In other words, for a large system, it is advantageous to use a larger number of swap moves on a single CPU than a smaller number of swap moves on many CPUs, at least using our current implementation of the algorithm on the LAMMPS package.

\subsection{Future directions}

In order to improve the efficiency of the hybrid method, in particular in the parallel case, several future directions are possible because some improvements could be made to our current implementation of the scheme in the LAMMPS package. One possibility would be to use a separate serial architecture for performing swap moves, while running the MD blocks in parallel. The SMC block would be performed on one processor only, avoiding costly inter-processor communication. In this case the MD blocks would be more efficient in parallel than in serial and the efficiency of the SMC blocks would be the same, meaning that overall this implementation should be faster. However, this method requires copying data from the LAMMPS parallel architecture and building neighbor lists from scratch before every SWAP block. Then, at the end of SMC blocks the new particle sizes would be sent back to each processor and the neighbour lists and parallel architecture updated again. It may be that these extra calculations will have a strong effect on the computational efficiency.

Another way to improve the speedup and circumvent the issues encountered while dealing with LAMMPS architecture would be to write a handmade molecular dynamics code that could be more versatile, and optimized for hybrid simulations. The MD part of the code in this case would be designed to run in parallel and integrate efficiently with a completely serial SMC routine.

\section{Continuous time swap MD algorithm}

\label{sec:continuous}

\subsection{Equations of motion}
\label{subsec:ham}

In this section, we introduce an algorithm that includes the physics of swap MC moves in a fully continuous time MD framework. The swap MC algorithm used in the context of supercooled liquids~\cite{GP01,NBC17} uses swap moves where the diameter of pairs of particles is exchanged, which leaves the particle size distribution fixed. In older versions of the swap MC algorithm~\cite{oldfrenkel}, particle diameters were exchanged with an external bath in a semi-grand canonical ensemble. This ensemble is conveniently used to describe theoretically~\cite{glandt} and numerically~\cite{glandt2} mixtures with a continuous size polydispersity. In this approach, the particle diameters are considered as fluctuating variables along with the particle positions. The diameters are constrained by an external potential (a chemical potential), and the particle size distribution becomes the result of the equilibrium sampling. The approach used in this section is a continuous time version of this idea. A zero-temperature version of the algorithm is discussed in Refs.~\cite{lernerprx,lernerquick}, which study the nature of energy minima generated by the Hamiltonian shown below in Eq.~(\ref{eq:ham}).

To study instead the finite temperature version of this approach, we introduce a generalised Hamiltonian where the diameters of particles are considered as dynamical variables, alongside their positions. For a system of $N$ particles, given the 3$N$ particle coordinates ${\bf r}^N \equiv \{{\bf r}_1,{\bf r}_2, \ldots ,{\bf r}_N\}$, their 3$N$ conjugate momenta ${\bf p}_{\bf r}^N \equiv \{{\bf p}_{1,{\bf r}}, {\bf p}_{2,{\bf r}}, \ldots , {\bf p}_{N,{\bf r}}\}$, the $N$ particle diameters $\sigma ^N\equiv \{\sigma_1, \sigma_2, \ldots ,\sigma_N\}$ and their $N$ conjugate momenta $p_\sigma^N \equiv \{p_{1, \sigma}, p_{2,\sigma}\ldots , p_{N,\sigma}\}$, we define the Hamiltonian
\begin{eqnarray}
\label{eq:ham}
H({\bf{r}}^N,{\bf{p_r}}^N,\sigma^N,p_\sigma^N)&=&\sum_{i}\frac{{\bf{p}}_{i,r}^2}{2m} + U({\bf{r}}^N,\sigma^N) \nonumber \\
                    & &  +\sum_{i}\frac{{p}_{i,\sigma}^2}{2M} +V(\sigma^N),
\end{eqnarray}
where $m$ is the mass conjugate to position momenta $p_r$, and $M$ is the mass conjugate to diameter momenta $p_\sigma$. The potential energy due to inter-particle interactions is given by $U({\bf{r}}^N,\sigma^N)$, which can be an ordinary pair potential. Each particle is additionally subject to a potential $v(\sigma_i)$ that constrains its diameter $\sigma_i$, so that $V(\sigma^N) = \sum_i v(\sigma_i)$ in Eq.~(\ref{eq:ham}). Examples of this potential will be given below.

The equations of motion follow from Hamilton's equations, and read
\begin{eqnarray}
\frac{\mathrm{d}{\bf p}_{i,{\bf r}}}{\mathrm{d}t} &=& -\frac{\partial H}{\partial {\bf r}_i}=-\frac{\partial U({\bf{r}}^N,\sigma^N)}{\partial {\bf r}_i}~~, \\
\frac{\mathrm{d}p_{i,\sigma}}{\mathrm{d}t} &=&-\frac{\partial H}{\partial \sigma_i}=-\frac{\partial [U({\bf{r}}^N,\sigma^N)+V(\sigma^N)]}{\partial \sigma_i}~~, \\
\frac{\mathrm{d}{\bf r}_i}{\mathrm{d}t}&=&\frac{\partial H}{\partial {\bf p}_{i,r}}=\frac{{\bf p}_{i,r}}{m}~~,\\
\frac{\mathrm{d}\sigma_i}{\mathrm{d}t} &=& \frac{\partial H}{\partial p_{i,\sigma}}=\frac{p_{i,\sigma}}{M}~~.
\label{eq:eom}
\end{eqnarray}

Similarly to standard molecular dynamics, we discretise in time these equations of motion and obtain an enlarged version of the standard velocity-Verlet algorithm. We solve the equations of motion using this algorithm with a time discretization $dt =0.001$. As a result, we obtain trajectories for the particles coordinates and diameter. In the following, we simulate systems of $N = 500$ particles at number density $N / L^3=1.0$ in canonical ensemble $NVT$ in cubic box with periodic boundary conditions.

We consider two temperatures $T_r$ and $T_\sigma$ related to particle translational momenta and diameter momenta, respectively. They are defined as
\begin{eqnarray}
T_r &=& \frac{1}{3N}\sum_{i} \frac{{\bf{p}}_{i,r}^2}{2m_i}~~,\\
T_\sigma &=&\frac{1}{N}\sum_{i}\frac{{p}_{i,\sigma}^2}{2M_i}~~.
\end{eqnarray}
The temperatures are kept constant and equal, $T_r=T_\sigma$, during the simulations using a Berendsen thermostat~\cite{berendsen} with coupling time constant $\tau=5.0$. In the following, we refer to the temperature simply as $T$. The reduced units are defined exactly as in Sec.~\ref{subsec:model}.

\subsection{Microscopic model}

\label{subsec:continuousmodel}

A glass-forming model is typically defined by the interactions between the particles and their size dispersity. We model the interaction between two particles $i$ and $j$ by the pair potential defined in Eq.~(\ref{eq:pot}). In the continuous method, we cannot use the same nonadditive cross diameter rule as presented in Sec.~\ref{subsec:model} because its derivative is not continuous. As a first step, we have simulated an additive rule for the diameters but we could easily generalise the nonadditive rule replacing the absolute value $|\sigma_i - \sigma_j|$ by a smooth function with equivalent properties, such as for instance $[1-\exp(-(\sigma_i - \sigma_j)^2)]$.

We focus on continuously polydisperse systems characterized by their diameter distribution, $P(\sigma)$. Contrary to the hybrid and swap MC methods in which particles exchange their diameters leaving the global distribution $P(\sigma)$ unaffected, the present method does not directly impose the diameter distribution $P(\sigma)$. Instead, one must impose an external potential for the diameters, $v(\sigma)$, to constrain the fluctuations of the diameters $\sigma^N$ and the particle size distribution is obtained as the result of the equilibrium simulations. This is a major difficulty if one wants to perform simulations at a series of state points, since $P(\sigma)$ would evolve if $v(\sigma)$ is left unchanged. Therefore, this approach needs an additional iteration step where the potential $v(\sigma)$ is adjusted at each state point in order to keep $P(\sigma)$ constant. This additional step becomes time consuming at low temperatures, where the equilibration of the system is slow and controls in particular the convergence of the distribution $P(\sigma)$ itself.

\begin{figure}
\centering
\includegraphics[width=7.5cm ]{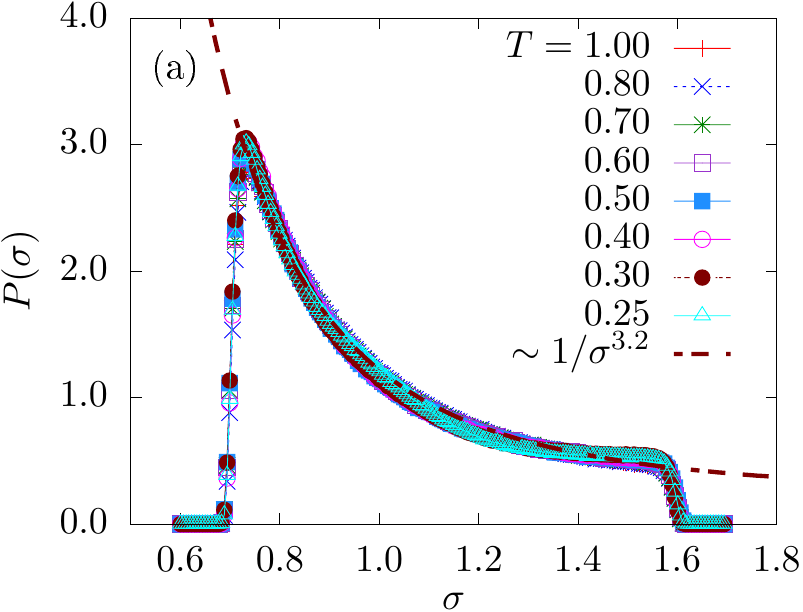}
\includegraphics[width=7.5cm]{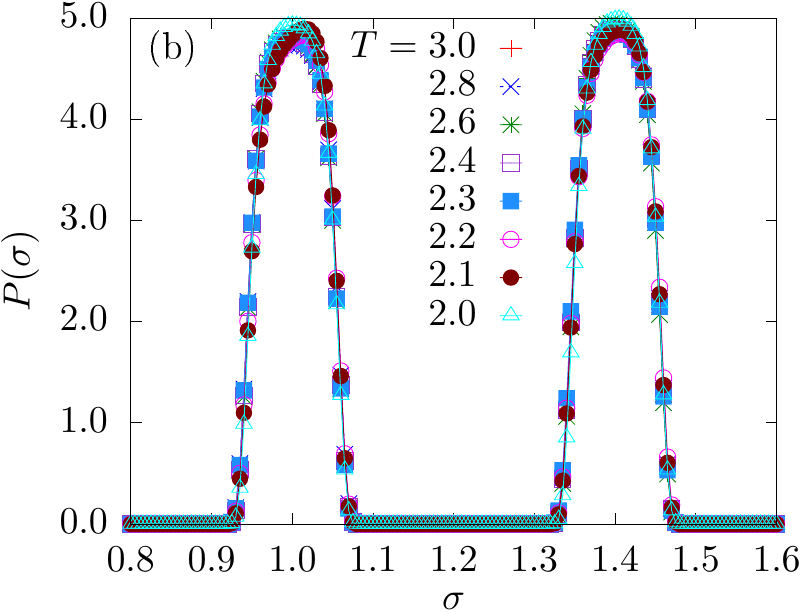}
\includegraphics[width=7.5cm]{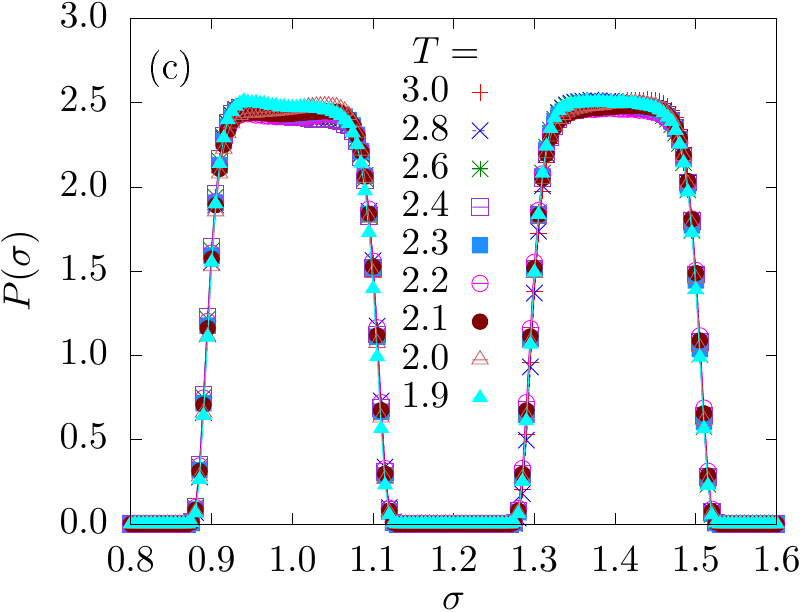}
\caption{Probability distribution of diameters $P(\sigma)$ measured in equilibrium at different temperatures $T$ for: (a) the continuous polydisperse system with $P(\sigma)\approx 1/\sigma^{3.2}$; binary systems with uniform distribution of width $\Delta \sigma = 0.1$ (b) and $\Delta \sigma = 0.2$ (c).}
\label{fig:dist_poly}
\end{figure}

We simulate two classes of systems which were shown to be structurally stable against crystallization at low temperature using swap MC. The first system is analogous to the continuously polydisperse one presented in Sec.~\ref{subsec:model}, and is characterized by $P(\sigma)\sim 1/\sigma^{3.2}$ in a finite range [$\sigma_m$, $\sigma_M$]. In order to obtain this diameter distribution at equilibrium, we design the diameter potential $v(\sigma)$ as follows. The hard boundaries of the distribution at $\sigma_m$ and $\sigma_M$ are imposed by two very steep exponential functions. To generate a power law distribution $P(\sigma)\sim 1/ \sigma^{3.2}$ in between, we employ a smooth power law form. The diameter potential used to enforce this distribution thus reads
\begin{eqnarray}
v(\sigma)  &= &  \mathrm{exp}[A(- \lambda_1 \sigma + \sigma_m)] + \mathrm{exp}[A( \lambda_2\sigma - \sigma_M)] - D \sigma^n \nonumber \\
& = & v_{\sigma}(A,\lambda_1,\lambda_2,D,n)~~, \label{eq:potpoly}
\end{eqnarray}
where the parameters ($A, \lambda_1, \lambda_2, D, n$) need to be tuned at each temperature in order to obtain the desired size distribution in equilibrium. More quantitative details on this procedure are given in Appendix~\ref{append:pot}, where all simulated parameters are tabulated. We show in Fig.~\ref{fig:dist_poly}a the measured probability distribution function at equilibrium across a range of temperatures. Therefore, we have successfully designed a diameter potential that imposes a constant diameter distribution that resembles the one studied in Sec.~\ref{sec:hybrid} with the hybrid method.

The second type of glass-forming model we study is a continuously polydisperse version of a discrete binary mixture, using a 50:50 mixture of particles with a typical size ratio $\sigma_B/\sigma_A=1.4$. In this model, the original delta peaks at $\sigma_A$ and $\sigma_B$ in the distribution $P(\sigma)$ are broadened uniformly over a typical width $\Delta \sigma$. We have considered two such binary systems of width $\Delta \sigma = 0.1$ and $\Delta \sigma = 0.2$. These diameter distributions are again designed by using the same functional form $v_\sigma$ as in Eq.~(\ref{eq:potpoly}) but using two distinct types of particles. This approach means we must now adjust 10 independent parameters at each simulated temperature. We show in Fig.~\ref{fig:dist_poly}b-c the measured probability distribution functions measured at equilibrium across a broad range of temperatures for the two systems considered. The quantitative details about the parameters used in these simulations are also tabulated in Appendix~\ref{append:pot}.

We demonstrate in Fig.~\ref{fig:dist_poly} that we are successful in designing diameter potentials and sets of parameters which produce a desired diameter distribution $P(\sigma)$ across different temperatures. This method, however, is relatively cumbersome. At each temperature, one has to make many trials in order to find the parameters for the diameter potential that yields the desired probability distribution at equilibrium. As temperature decreases, relaxation times increase and the trial and error procedure becomes increasingly costly in terms of CPU time. In our effort to design new algorithms and methods to simulate supercooled liquids at ever lower temperature, this method therefore does not necessarily appear as the most efficient one, as it introduces the need to perform a large number of runs to prepare the system before making any measurement. Of course, the temperature evolution of the potential $v(\sigma)$ is very smooth, and thus training at high temperatures and some educated guesses help converge that procedure faster.

\subsection{Structural instability and crystallization}
\label{subsec:unstable}

Crystallisation, fractionation and ordering are problems that need to be faced when dealing with supercooled liquids. To push the swap MC method to its maximal efficiency, new models of supercooled liquids were developed that better resist ordering and are therefore better glass-formers. Recent investigations have demonstrated that swap MC is able to crystallise polydisperse models of hard spheres relatively easily~\cite{BBBT18,truskett,engel}, whereas the ordinary dynamics would only allow one to probe the metastable fluid.

\begin{figure}
\centering
\includegraphics[scale =1 ]{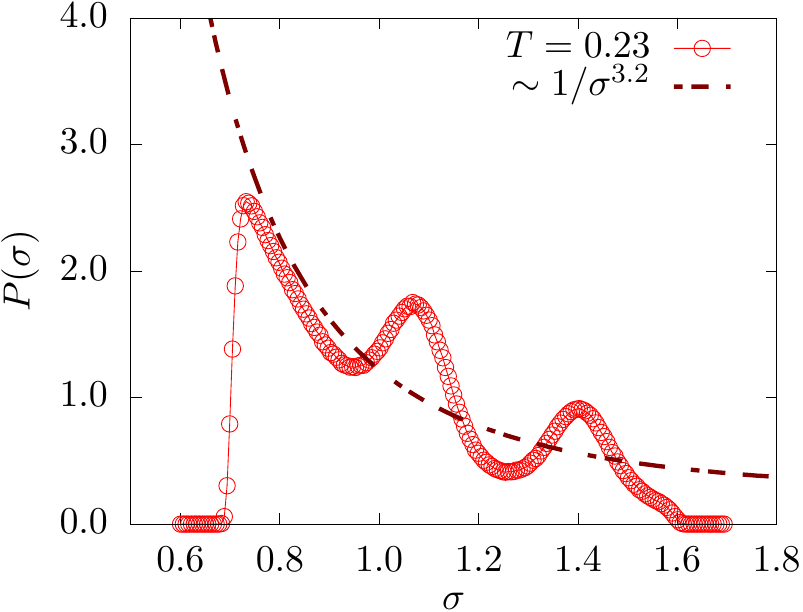}
\caption{Probability distribution $P(\sigma)$ of diameters $\sigma$, measured  for continuous polydisperse system at $T=0.23$, where phase separation and crystallisation is observed. The energy cost in diameter space due to the distortion of the particle size distribution is more than compensated by an ordering in position space.}
\label{fig:cryst}
\end{figure}

We expect that the hybrid method and swap MC behave similarly with respect to crystallisation, but we find that the fully continuous version shows qualitatively distinct behaviour, as we now explain. In Fig.~\ref{fig:dist_poly}a, we show that a continuous polydispersity can be easily maintained down to $T=0.25$ by a proper choice of the potential $v(\sigma)$. If we use this insight to attempt thermalising the system at $T<0.25$ we observe that the system becomes unstable. An example is shown in Fig.~\ref{fig:cryst} which shows the measured $P(\sigma)$ for $T=0.23$, compared to the functional form $\sigma^{-3.2}$ observed at higher temperatures. It is clear that the shape of the particle size distribution is now completely different since it develops peaks near $\sigma\sim 1.0$ and $\sigma \sim 1.4$. Simultaneously, direct visualisation reveals that the system has partially crystallised and phase separated between large and small particles.

The physical interpretation is that the system is distorting the particle size distribution (thus paying an energetic cost in diameter space due to the potential $v(\sigma)$) in order to gain free energy by ordering the system in position space. Such instability is typically not observed using hybrid and swap MC algorithm because the particle size distribution is by construction not allowed to vary over the course of a simulation. Of course, in the large system size limit, the phase separation and crystallisation reported in Fig.~\ref{fig:cryst} should also occur when swap MC is used, because concentration fluctuations would occur. These fluctuations are presumably too slow to lead to crystallisation in hybrid and swap MC approaches.

We conclude, therefore, that the type of semi-grand canonical simulation that we perform when employing the continuous time swap algorithm may unfortunately accelerate the crystallisation of the system. To cure this problem, new models should be developed that are even more robust against ordering and could then be simulated using the continuous time swap algorithm. For instance, one could attempt to use a finite nonadditivity to the pair interaction as a first step in this direction.

\subsection{Choice of the diameter mass}

In standard molecular dynamics the diameter of the particles is constant. This corresponds to the limit of an infinite diameter mass $M$ in the continuous method described by Eq.~(\ref{eq:eom}). As the diameter mass decreases from $M=\infty$, the diameters become dynamical variables and start to vary and influence the structural relaxation.

\begin{figure}
\centering
\includegraphics[scale =1]{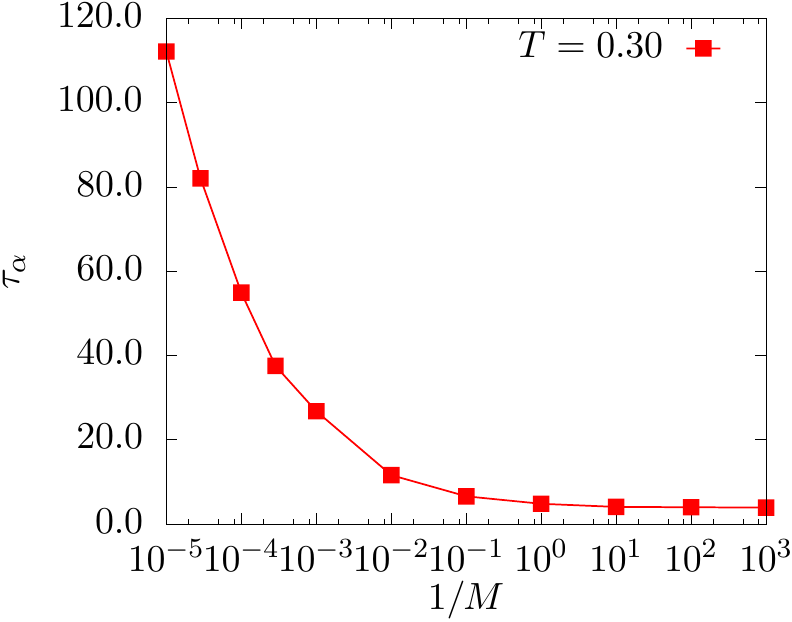}
\caption{Dependence of the equilibrium relaxation time $\ta$ with inverse diameter mass $1/M$ for continuous polydisperse liquids $P(\sigma)\sim 1/\sigma^{3.2}$ at temperature $T=0.3$.}
\label{fig:massDepend}
\end{figure}

We look for the diameter mass that optimizes the continuous method. To do so, we compute the relaxation time $\ta$ of a liquid as a function of the diameter mass $M$. The relaxation time is computed as in Sec.~\ref{subsec:physeff}, taking Eq.~(\ref{eq:self}) at wavevectors of magnitude $k=6.7$. We report in Fig.~\ref{fig:massDepend} the measured relaxation time $\ta$ as a function of inverse diameter mass $1/M$ in the continuously polydisperse system $P(\sigma) \sim  1/\sigma^{3.2}$, at a fixed temperature $T=0.30$. 

The qualitative behavior of $\ta$ in Fig.~\ref{fig:massDepend} is qualitatively similar to the one in Fig.~\ref{fig:rsvar}. The parameter $1/M$ plays a role similar to the swap density $\rs$ or the probability $p$ of particle-swap moves in the hybrid and swap MC algorithms, respectively. When they increase, the typical timescale for the diameter dynamics decreases, which speeds up the physical relaxation of liquids. We observe a clear decrease in the structural relaxation time as the mass $M$ of diameters decreases, starting from a very large value $M = 10^5$. Around $M=1$, the relaxation time reaches a plateau, and decreasing further the diameter mass $M$ does not speed up the structural relaxation of the liquid.

While any choice $M < 1$ minimizes the time needed to relax the liquid, a very small diameter mass is not suitable. Indeed, when $M$ is too small, large variations of the diameters occur on very short time scales, which requires a very small integration time step $dt$. This effectively increases the CPU time of the simulations, which is undesired. In the following, we choose $M=1$ as the optimal compromise between physical speedup and computational efficiency.

\subsection{Physical Efficiency }

We now compare the relaxation dynamics of the three glass-forming models presented in Sec.~\ref{subsec:continuousmodel} when simulated both with the continuous swap method and standard MD simulations. We first have to determine iteratively the correct parameters for the diameter potential at each temperature, as described above. Then, we run simulations with the continuous method at a temperature $T$ to obtain equilibrium configurations. We also measure the equilibrium relaxation time of the system using the continuous time swap algorithm. Finally, the equilibrated configurations are taken as initial conditions for standard MD simulations, during which the relaxation time is measured. By construction, then, the MD simulations run the dynamics for the same particle size distribution as the continuous time swap algorithm.
These configurations will also serve as starting points for hybrid MD/MC simulations, to be discussed below in Sec.~\ref{sub:comparison-hybrid}.

\begin{figure}
\centering
\includegraphics[width=7.5cm]{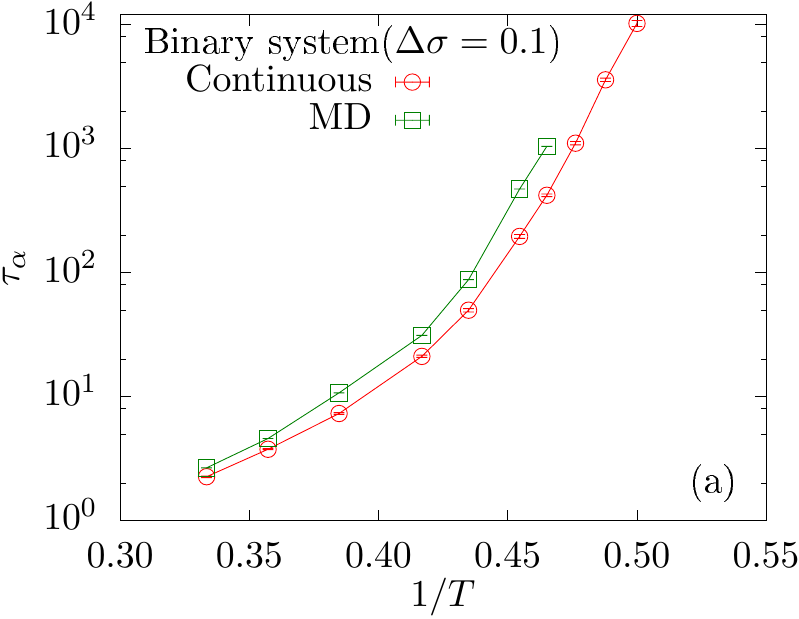}
\includegraphics[width=7.5cm]{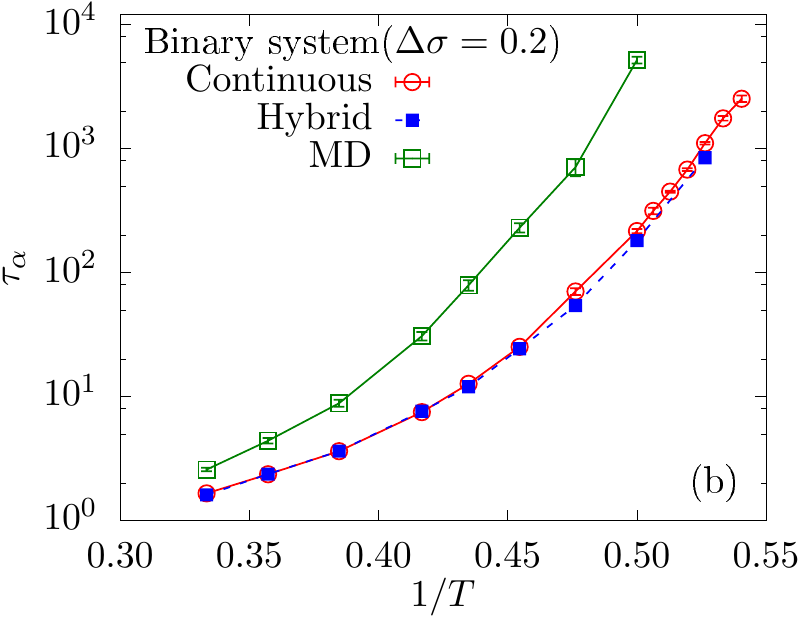}
\includegraphics[width=7.5cm]{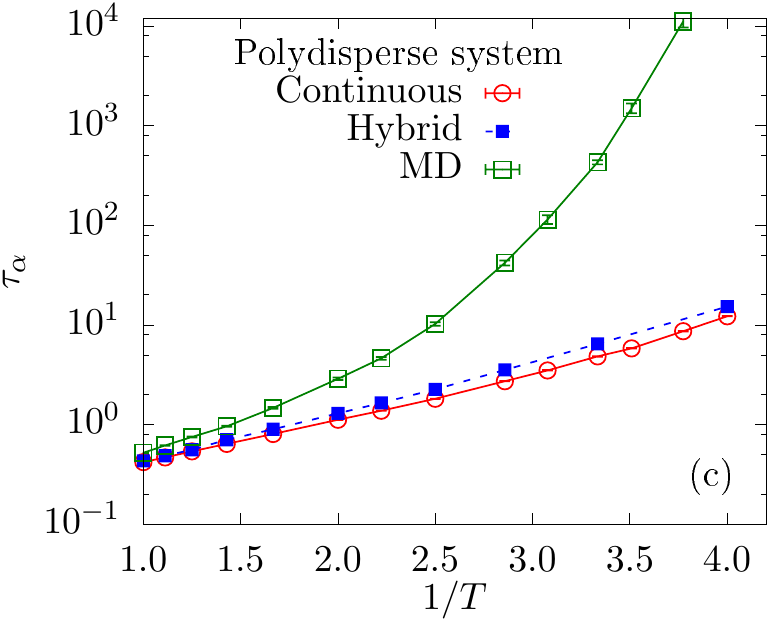}
\caption{Relaxation time as a function of inverse temperature measured in: (a) binary system with $\Delta \sigma =0.1$, (b) $\Delta \sigma =0.2$, and (c) continuous polydisperse system. Relaxation times have been computed using three different methods continuous time swap method, hybrid method and standard MD.}
\label{fig:physSwapMD}
\end{figure}

The results for the equilibrium relaxation time $\ta$ of the three models and three numerical algorithms are reported in Fig.~\ref{fig:physSwapMD}. The binary systems with $\Delta \sigma =0.1$ and $\Delta \sigma =0.2$ can be simulated down to quite low temperature without crystallizing. In both cases, the efficiency of the continuous time swap method over MD simulations is temperature dependent, with an efficiency increasing as temperature decreases. The speedup in thermalization offered by the continuous method depends on the width $\Delta \sigma$ accessible to diameters. Larger variations in the particles' diameters are expected to ease even more the structural relaxation of the liquid. When $\Delta \sigma = 0.1$, diameters are more constrained than when $\Delta \sigma =0.2$. The dynamical gain observed in Fig.~\ref{fig:physSwapMD}(a) is about one order of magnitude in relaxation time for $\Delta \sigma =0.1$, while for $\Delta \sigma =0.2$, extrapolation of the data presented in Fig.~\ref{fig:physSwapMD}b suggest that the continuous time swap method can more easily achieve thermalization in a region inaccessible to MD simulations, with a speedup estimated at about 2 orders of magnitude. Clearly these two binary systems do not yield as large a speedup as fully polydisperse models~\cite{NBC17}, and as a result are less prone to crystallisation.  

In the case of the continuous polydisperse model with diameter distribution $P(\sigma) \sim 1/\sigma^{3.2}$, shown in Fig.~\ref{fig:physSwapMD}c, the dynamical gain with the continuous method is even greater, similarly to what was measured with the swap MC algorithm\cite{NBC17}. While the dynamical gain is very interesting with this model, the continuous time method cannot simulate supercooled liquids at temperatures lower than $T<0.25$, the last point studied, because of structural instability discussed in Sec.~\ref{subsec:unstable}, and we can thus not benefit from the efficiency of the swap algorithm as much as when the hybrid method is used.

\subsection{Comparison with the hybrid method}
\label{sub:comparison-hybrid}

In this section, we compare the physical efficiency of the continuous time swap algorithm with the hybrid method. To that effect, we use configurations equilibrated with the continuous method as initial condition for hybrid simulations using parameters ($\ns = 10$, $\tmd=0.1$), and measure the relaxation time of the liquid. Results for the binary system with $\Delta \sigma =0.2$ and the continuous polydisperse system with $P(\sigma) \sim 1/\sigma^{3.2}$ are presented in Fig.~\ref{fig:physSwapMD}b-c. We observe that both methods give physical relaxation times that are extremely close to one another, and have a very similar temperature dependence. Note that we did not tune the parameters of each method in order to obtain the exact same relaxation times, but rather used each technique with its own set of optimal parameters. The agreement between the two methods suggests that the continuous time swap method, once optimised, captures the same physics as the other swap algorithms (hybrid MC/MD and pure MC). Overall, we conclude that all three algorithms have the same efficiency in terms of speedup of the structural relaxation.

\begin{figure}
\centering
\includegraphics[scale =1]{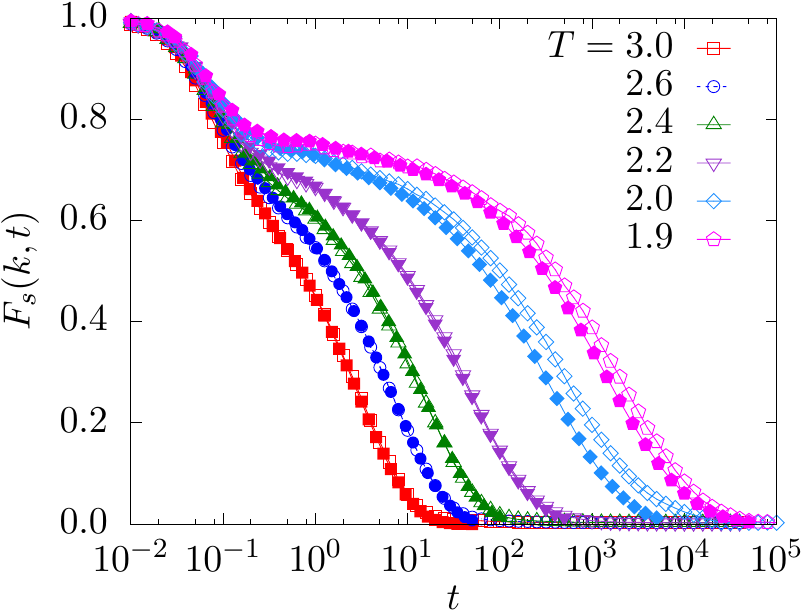}
\caption{Self-intermediate scattering function $F_s(k,t)$ calculated for the binary system with $\Delta \sigma =0.2$ using the continuous time (open symbols) swap and hybrid (filled symbols) algorithms at different temperatures.}
\label{fig:scatFun}
\end{figure}

Finally, we also plot the self-intermediate scattering functions measured at different temperatures in hybrid and continuous time methods in Fig.~\ref{fig:scatFun}. At each temperature, the curves corresponding to the two methods have the same time dependence. Both the relaxation dynamics at long times and microscopic dynamics are very similar. This implies that the two methods are equivalent as far as the relaxation of the liquid is concerned. The different nature of the microscopic rules for the dynamics do not matter. Performing discrete particle-swap moves or continuously modifying the diameters of particles has no influence on the physical relaxation of supercooled liquids at long times. What matters, eventually, is the strong coupling between diameter and position degrees of freedom that relax in a strongly correlated manner~\cite{NBC17}, the diameter fluctuations allowing the system to efficiently relax the positional degrees of freedom even in a temperature regime where the physical dynamics is extremely slow.

\subsection{Computational performance}

The continuous time swap algorithm runs similarly to conventional MD simulations, with the difference that particles have one more degree of freedom (the diameter) in addition to the positions. This implies that running this algorithm is essentially as costly in terms of CPU time as running a conventional MD simulations. But the speedup offered in terms of structural relaxation time is the same as with swap MC method. This method thus offers a valuable alternative to swap MC, especially for users that are not familiar to Monte Carlo simulations.

Attempts to parallelize the hybrid method did not bring significant improvements in terms of CPU time. This was due to the difficulty to parallelize efficiently the Monte Carlo blocks present in the hybrid method. The continuous time swap method offers the same advantages as standard molecular dynamics simulations in terms of parallelisation. Since the dynamics is continuous and deterministic, one can in principle implement this method to run it efficiently on several processors. The CPU time needed to run simulations of the same MD length is expected to scale with the number of particles per processor, as discussed in Sec.~\ref{subsec:parallel}.

\section{Discussion and perspectives}
\label{discussion}

In this work, we provided two distinct generalisations of the swap Monte Carlo algorithm that has recently proved extremely successful in producing equilibrium configurations of supercooled liquids at very low temperatures. Both algorithms combine the idea of particle swaps with conventional Molecular Dynamics techniques. In the first version, we simply alternate periods of conventional MD with periods of swap MC moves, while in the second we solve Hamilton's equations of motion for both positions and diameters simultaneously, in a fully continuous time MD scheme.

After an adequate optimisation of all simulation parameters involve in each three swap-like algorithms, we find that the three algorithms provide a very similar (and quite impressive in some cases) speedup of equilibration, which suggests that the same physics is at play in the three cases. Namely, the addition of diameter fluctuations strongly couples to positional degrees of freedom to relax the structure of the supercooled liquid. The equivalence between the three algorithms even extends to time correlation functions.

Our general conclusion is that all three algorithms can be equivalently used to produce low-temperature equilibrium configurations for model glass-formers, and which algorithm should be preferred depends is firstly a matter of personal convenience. The hybrid and swap MC are very close to one another in spirit and performances, and the implementation into the LAMMPS software of the hybrid method makes it user-friendly in case a different model needs to be studied. Regarding the continuous time swap algorithm, it is promising since it combines the efficiency of the swap MC to the simplicity of the MD technique, with great potential if large systems need to be studied. However, the iterative determination of the diameter potential makes it more cumbersome to use, and one must find ways to prevent the ordering that the semi-grand canonical ensemble seems to facilitate.
In future work, it would therefore be interesting to develop more robust glass-forming models that can resist the crystallisation and phase separation observed when the particle size distribution is not conserved by the dynamics.

\bibliography{biblio}

\appendix

\section{Considerations for implementing the hybrid MD/MC swap algorithm in LAMMPS}

\label{sec:lammps}

In this section of the appendix we give some details about how the hybrid molecular dynamics/particle-swap Monte Carlo method is implemented in the LAMMPS package. We provide an outline of how several problems were overcome.

\subsection{Handling continuously polydisperse systems in LAMMPS}

In LAMMPS, each particle has a type and all particles of a given type share the
same values for certain properties. One of these shared properties is their size $\sigma$. This means that to simulate a system of $N$ particles with continuous polydispersity, $N$ different types of particle are needed. Defining $N$  types of particles in LAMMPS would be cumbersome, especially if we wish to simulate large numbers of particles.

To overcome this problem, we decided to define only one particle type, and to store the diameters of particles in a type-independent
property. We used the existing \texttt{charge} property to store the diameter of each particle. We also created a modified version of \texttt{pair\_style lj} called \texttt{pair\_style lj\_poly} that
uses the \texttt{charge} in place of particle size when calculating the pair
interaction energy.

\subsection{Avoiding neigbour list rebuilds after every swap move}

To keep the neighbour list of particle $i$ as short as possible, LAMMPS takes
the size of particle $i$ into account when generating its neighbour list. This
means that if the size of $i$ should change (for example during a swap move),
the neighbour list is incorrect and must be recalculated. Given that we
typically attempt $N$ swap moves after every MD step, recalculating neighbour
lists this frequently would overwhelm any computational time gained by using
the swap algorithm.

To reduce this computational burden, we calculate the neighbour list for
particle $i$ as if it had the largest size in the particle size distribution.
This means that after a successful swap move, the neighbour list will still be
valid. This modification comes at the price of longer neighbour lists, but the
increase in the time to calculate pair interaction energies and to update a
particle's neighbour list is offset by not having to recalculate the neighbour
list after every swap move. We note that this means the simulation time for
systems of particles with pair interactions that have short cutoffs will be
significantly faster.

\subsection{Full and half neighbour lists}

The neighbour lists required by Molecular Dynamics and Monte Carlo simulations
are different. Due to the nature of the energy and force calculations being
carried out at each step, Molecular Dynamics simulations require that a
pair of particles appears once in the neighbour lists: that is if $j$ is in the
neighbour list of $i$ then $i$ is not in the neighbour list of $j$. In a Monte
Carlo simulation, particle $i$ must know about all the particles it could
interact with: $i$ would appear in the neighbour list of $j$ and $j$ in that of
$i$. In practice this means that the neighbour lists required for Monte Carlo
simulations are twice the size of those for Molecular Dynamics. In LAMMPS, these are referred to as \texttt{full} and \texttt{half} neighbour lists. Since the energy and force
calculations take up the bulk of computational time, we wish to avoid
maintaining only \texttt{full} neighbour lists and thus doubling the length
of the force calculations performed during each Molecular Dynamics move. The
alternative solution of maintaining only \texttt{half} neighbour lists and performing a
sum over all particles for the energy calculations during Monte Carlo moves
is even less desirable.

Thankfully LAMMPS has a method for updating \texttt{full} and \texttt{half}
neighbour lists together at the same time - the computational overhead to do this is
considerably less than that required for the two solutions described above.
The class \texttt{pair\_lj\_poly} must be written to ensure that the correct neighbour
list is used in each case: \texttt{full} for interaction energies and \texttt{half} for force
calculations.

\subsection{Triggering blocks of swap moves}

Due to some technical details about how blocks of swap moves are triggered
during a LAMMPS simulation we had to modify the \texttt{run} function in the LAMMPS
\texttt{verlet} class. The swap moves are triggered within a \texttt{modify->pre\_exchange()}
command and the position of this command in the \texttt{run} function means that
neighbour lists are unnecessarily calculated every time a block of swap
moves is attempted. We moved the position of the \texttt{modify->pre\_exchange()}
command within the \texttt{run} function to prevent this.

\subsection{The hybrid method and parallelisation}

The final issue, which remains only partially resolved, arises due to
integrating a serial simulation method (swap Monte Carlo) into a parallelized
one (Molecular Dynamics, as performed by LAMMPS). This issue is caused by the
particular way that LAMMPS implements parallel computation and could be avoided
if custom Molecular Dynamics code was used. If this was done, the theoretical
maximum computational efficiency for the parallelised hybrid method would be
 achieved.

LAMMPS spatially parallelises the system, meaning that a processor has
responsibility for a sub-box of the simulation box. A processor must keep
track of particles on neighbouring processors that particles it has
responsibility for may interact with. This means that during Molecular Dynamics
bouts of inter-processor communication must be carried out with roughly the
same frequency as the neighbour lists are rebuilt. Due to the non-local nature
of the changes that take place during swap moves and the way that LAMMPS keeps track of particle identities, this inter-processor
communication must be carried out much more frequently when attempting swap moves. We have tried to minimise
it as much as possible, but it is impossible to
eliminate without more serious modifications to LAMMPS. Unfortunately, these
communications are sufficiently costly that our implementation of the hybrid
method does not scale as well as it could when run on multiple processors.

\section{Designing diameter potentials in the continuous time swap algorithm}
\label{append:pot}

\subsection{Continuously polydisperse model}

In this method, in the absence of a diameter potential, $i.e.$ when $v_\sigma=0$, the particles sizes will all shrink to zero to minimize the potential energy. Therefore we must perform our simulations for a finite diameter potential to constrain the diameter sizes for a desired range and distribution. We employ Eq.~(\ref{eq:potpoly}) as defined in Sec.~\ref{subsec:continuousmodel} to generate continuously polydisperse systems with a size distribution $P(\sigma)\sim 1/\sigma^{3.2}$. In this equation, two exponential functions create the steep walls (the steepness is  determined by the parameter $A$, here $A=100.0$) at minimum $\sigma_m$ and maximum diameters $\sigma_M$. To generate the desired particle size distribution between $[\sigma_m, \sigma _M]$, we employ a power law form with proper combination of parameters $n$ and $D$. The power $n$ decides the nature of the distribution, while the prefactor $D$ set an energy scale in diameter space (and hence is $T$-dependent).

We start the process of tuning the parameters of diameter potential at some initial temperature. We first obtain $n=2.6$ and  $D=14.46$ at $T=1.0$. We know that for a given size distribution, the pair potential energy increases as $T$ increases. So if we fix these potential parameters $n$ and $D$ and investigate a higher $T$, the kinetic energy will not suffice to sample enough of the large particles and we need to increase the parameter $D$ to reobtain the correct distribution. Similarly we decrease the parameter $D$ as we decrease $T$.  Also, we notice that after fixing the parameters $A$, $n$ and $D$, then while going from high to low $T$, the particle size distribution becomes systematically narrower and therefore we need to choose two more parameters, $\lambda_1$ and $\lambda_2$, to maintain the correct width of the size distribution.

\begin{figure}
\centering
\includegraphics[scale = 1]{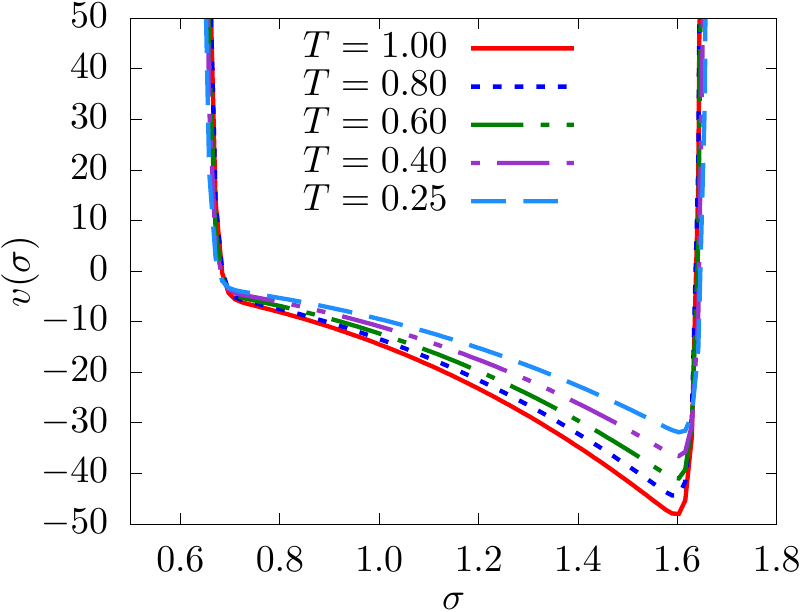}
\caption{Diameter potential at different $T$ for suitable set of parameters to maintain the desirable distribution of particles with average diameters $\sigma\approx1.0$ and polydispersity of $\approx 24\%$. }
\label{fig:pot_poly}
\end{figure}

\begin{table}[b]
\centering
\begin{tabular}{c c c c }
\hline
\hline
 $T$ & $D$ &$\lambda_1$ & $\lambda_2$ \\
\hline
\hline
 1.00 & 14.4600  & 1.000 & 1.0000 \\
 0.90 & 13.9400  & 1.000 & 1.0000 \\
 0.80 & 13.3900  & 1.000 & 1.0000 \\
 0.70 & 12.8830  & 1.004 & 0.9995 \\
 0.60 & 12.2998  & 1.006 & 0.9980 \\
 0.50 & 11.6840  & 1.009 & 0.9950 \\
 0.45 & 11.3139  & 1.010 & 0.9950 \\
 0.40 & 10.8932  & 1.010 & 0.9950 \\
 0.35 & 10.5277  & 1.013 & 0.9950 \\
 0.30 & 10.0566  & 1.016 & 0.9950 \\
 0.25 & 9.44502  & 1.017 & 0.9920 \\
 \hline
 \hline
\end{tabular}
\caption{Parameters for internal potential $v(\sigma)$ to generate distribution of size of particles $P(\sigma) \sim 1/\sigma^{3.2}$}
\label{table:tabpoly}
\end{table}

Here we tune our potential parameters such that the average size $\overline{\sigma} \approx 1.0$ and the average polydispersity is $\approx 24 \%$. The resulting potential is continuous in its first and second derivative and is thus convenient for MD simulations. Representative potential $v(\sigma)$ are shown in Fig.~\ref{fig:pot_poly} and the corresponding values of the parameters at different temperatures are reported in Table~\ref{table:tabpoly}.

\subsection{Binary polydisperse model}

The second class of system that we consider is an equimolar mixture of particles having average size ratio 1.4 ($i.e.$, $\sigma_A/\sigma_B=1.4$) with uniform distributions centered around their respective average diameters $\sigma_A$ and $\sigma_B$, of width $\Delta\sigma=0.1$ and $\Delta\sigma=0.2$.

To generate the diameter potential we employ the same functional forms as in Eq.~(\ref{eq:potpoly}).
There are six terms in this diameter potential. Four exponential functions (with steepness parameter $A=100$) define the steep walls delimiting the range of the particle size distributions of width $\Delta \sigma$, and two power law functions with suitable power of $\sigma$ ($n_1=2.5$ and $n_2=2.4$) produce uniform distributions centered around $\sigma_A=1.0$ and $\sigma=1.4$.

In this case, the average diameter is therefore $\overline \sigma \approx 1.2$. For the case of $\Delta \sigma = 0.1$, the polydispersity for $A$-type particles is is $\approx 3.2\%$ and around $\sigma_B=1.4$ it is $\approx 2.2\%$. For the case of $\Delta \sigma = 0.2$, the polydispersity around $\sigma_A=1.0$ is $\approx 6\%$ and around $\sigma_B=1.4$ it is $\approx 4.2\%$. The parameters used at different temperatures are listed in  Table~\ref{tab:tabwid0.1}.


\begin{table}
\centering
\begin{tabular}{c c c c c}
\hline
\hline
 $T$ & $D_1$ & $D_2$ & $\lambda_1$ & $\lambda_2$ \\
\hline
\hline
3.000  & 68.7081 & 72.1031 & 0.996  & 1.003 \\
2.800  & 67.5904 & 71.1848 & 0.997  & 1.002\\
2.600  & 65.7046 & 69.3978 & 0.997  & 1.002\\
2.400  & 64.3770 & 68.3707 & 0.998  & 1.001\\
2.300  & 63.9399 & 67.8369 & 0.999  & 1.001\\
2.200  & 63.5409 & 67.3381 & 0.999  & 1.000\\
2.150  & 62.8426 & 66.8393 & 0.999  & 1.000\\
2.100  & 62.6431 & 66.6398 & 0.999  & 1.000\\
2.050  & 61.6456 & 66.2408 & 0.999  & 1.000\\
2.000  & 61.2466 & 65.9415 & 0.999  & 1.000\\
 \hline
 3.000 & 69.0051 & 72.8954 & 0.996  & 1.003\\
 2.800 & 67.5200 & 71.3107 & 0.996  & 1.002\\
 2.600 & 65.8695 & 69.5649 & 0.998  & 1.002\\
 2.400 & 64.5384 & 68.5352 & 0.999  & 1.001\\
 2.300 & 63.9399 & 67.8369 & 0.999  & 1.001\\
 2.200 & 63.2000 & 67.2000 & 1.000  & 1.000\\
 2.100 & 62.3200 & 66.5500 & 1.000  & 1.000\\
 2.000 & 61.2000 & 65.7000 & 1.000  & 1.000\\
 1.975 & 60.9500 & 65.4500 & 1.000  & 1.000\\
 1.950 & 60.6500 & 65.2000 & 1.000  & 1.000\\
 1.925 & 60.3000 & 64.9000 & 1.000  & 1.000\\
 1.900 & 60.0500 & 64.8000 & 1.000  & 1.000\\
 1.875 & 59.9000 & 64.6500 & 1.000  & 1.000\\
 1.850 & 59.5550 & 64.4560 & 1.000  & 1.000\\
\hline
\hline
\end{tabular} 
\caption{Parameters used to design the potential $v(\sigma)$ to generate a binary distribution of diameters with width $\Delta\sigma=0.1$ (top) and $\Delta \sigma = 0.2$ (bottom).}
\label{tab:tabwid0.1}
\end{table}

\end{document}